\documentstyle[prd,tighten,aps]{revtex}
\begin{document} 
\draft 
\twocolumn[\hsize\textwidth\columnwidth\hsize\csname
@twocolumnfalse\endcsname

\title{ Covariant path integral for chiral p-forms }
\author{Fernando P. Devecchi and Marc Henneaux}
\address{ Facult\'{e} des Sciences, 
ULB-Campus Plaine, C.P. 231, 1050  
Bruxelles, Belgium.} 
\maketitle 
\begin{abstract} 
The covariant path integral for chiral bosons obtained by McClain, Wu and Yu
is
generalized to chiral p-forms. In order to handle the reducibilty of the gauge
transformations 
associated with the chiral p-forms and with the new variables
(in infinite number) that must be added to eliminate the second class
constraints, the field-antifield 
formalism is used.
\end{abstract}
  
\pacs{11.10.Ef, 12.10.Gq, 04.65.+e}

\vskip2pc]

\section{introduction}

Chiral p-forms play a central role in supergravity and in string theory
\cite{Ma,Gre}. In particular, 
they contribute to the
`miraculous" cancellation of the gravitational anomaly in type-IIB
supergravity
or superstring theory, making these theories quantum-mechanically consistent.

The calculation of the gravitational anomaly for chiral p-forms was performed
first in\cite{Alv} without using a Lagrangian but by guessing suitable Feynman
rules that incorporate the chirality condition. A Lagrangian that leads
to the correct equations 
of motion for chiral p-forms was given later in \cite{Henteit1,Henteit2}
both in flat and in curved space-times. This Lagrangian generalizes to chiral 
p-forms the one constructed in\cite{Flo} for chiral bosons in two dimensions,
a model
that has been extensivelly analysed during the last years \cite{Giro}.
Using the Lagrangian of \cite{Henteit1} the authors of \cite{Nie} recalculated
the gravitational anomaly for chiral p-forms and found agreement with the work
of \cite{Alv} even though their Feynman rules turned out to be different.

One feature of the Lagrangian given in  \cite{Flo} for chiral bosons, and of
its
generalization given  in \cite{Henteit1} for chiral p-forms, is that it is not
manifestly
covariant. Furthermore, it leads to second class constraints in the
hamiltonian formalism,  which imply non
usual commutation relations between the field variables.

In order to cure these difficulties, an infinite number of auxiliary fields
were introduced in \cite{Mc}.
These auxiliary 
variables do not carry physical degrees of freedom of their own and enable one
to replace the second class  constraints enforcing the chirality condition by
an infinite number of first class ones, along the lines of \cite{Fad}
and \cite{Bat}. These first class 
constraints generate a new gauge freedom and by
fixing the gauge through canonical methods one falls back on the original
description of the chiral boson.

Using this new formulation, the authors of \cite{Mc} were able to derive a 
covariant path integral for chiral bosons and to show that it reproduces
the correct physical
amplitudes and anomalies.

The purpose of this paper is to generalize the path integral derivation
of \cite{Mc} for chiral p-forms. The procedure is not entirely trivial because
chiral p-forms have already a gauge invariance of their own - contrary to
chiral bosons -,  which is 
furthermore reducible. Moreover, it turns out that the new gauge invariance
associated with the 
infinite number of auxiliary fields added to achieve covariance,  mixes in a
non-trivial way with the
original invariance of 
the p-forms, leading to even
more reducibility.  This  requires the presence of further ghosts of ghosts.

The more convenient way to handle this problem is to follow the lines of the
field-antifield
formalism \cite{Batil,Hen,Gom} as we do here.

Our paper is organized as follows. First we reproduce the results of \cite{Mc}
through the antifield approach. We then go to the p-form case. We derive the
explicit form of the pure first class action, introducing an infinite
number of auxiliary
fields. This first class description is verified to be
physically  equivalent to the second class description of
\cite{Henteit1,Henteit2},
as in the chiral boson case\cite{Mc}. We then derive the solution of the master
equation and provide a gauge fixing fermion leading to the desired
manifestly covariant
path integral. We close our paper with some comments on the applications
of this work.

\section{Covariant path integral for a chiral boson}

\subsection{Classical analysis}

It was recognized in \cite{Mc}  that a 
(1+1)-dimensional
chiral boson (= chiral 0-form)
could be consistently formulated in terms of the action

\begin{equation}
{S_0} =\int d^2x \left(\sum ^{\infty}_{n=0}\pi _n\dot \phi_n-{\cal H}\right)+
\int d^2x \sum _{n=1}^{\infty}\lambda _n T_n \,\,\,\, ,\label{1}
\end{equation}
where ${\cal H}$ is the Hamiltonian density

\begin{equation}
{\cal H } = \frac{1}{2}\sum ^{\infty }_{n=0} ({\pi _n}^2+
{\phi _n'}^2) (-1)^n\,\,\, .\label{2}
\end{equation}                    
The $\phi_n$ ($\phi _n\equiv \phi _n(x_0, x_1)\equiv \phi _n(\tau , \sigma)$,
$\dot \phi _n\equiv 
\partial _0 \phi
_n \,\,\, \phi _n'\equiv \partial _1 \phi _n$) are an infinite collection of
scalar 
fields,  the $\pi_n$ are their
conjugate momenta. The $T_n$ constitute an infinite set of first class 
constraints and are explicitly given by

\begin{equation}
T_m=\pi_{m-1}-\phi _{m-1} '+\pi_m+\phi _m'\approx 0 \,\,\,\,\,\,\,  m\geq 1
 .\label{3} 
\end{equation}
One has for the equal time brackets
\begin{equation}
\left[ T_l(\sigma ),T_m (\sigma ')\right]=0 \label{4}
\end{equation}
as well as
\begin{equation}
\left[ { H} , T_m(\sigma )\right] =(-1)^{m+1}T_m' (\sigma )\,\,\,\, ,
 \,\, { H}
=\int d\sigma '
{\cal H} \,\,\,\,\, .\label{5}
\end{equation}   
The $\lambda _n$ are Lagrange multipliers for the constraints (\ref{3}).

The system  above was obtained in \cite{Mc} for the model of \cite{Flo}
by enlarging the  number of fields in such a way that the original second
class constraint 
$(\pi _0 -{\phi_0}'\approx 0)$, enforcing the chirality of $\phi _0$,
is replaced by a collection of constraints that are all first class. This can
be achieved along the systematic lines of \cite{Fad,Bat} and leads to
an infinite
tower of constraints (see also \cite{Wot} for a related discussion).
To  show the equivalence of (\ref{1}) with the Floreanini-Jackiw action
\cite{Flo}
($ S=\int d^2x(\phi _0' \phi _0- \phi _0' \phi _0'$)), one observes that
(\ref{1})
is invariant under the following gauge symmetries
generated by the first class constraints
(\ref{3}),   
\begin{mathletters} 
\label{6}
\begin{eqnarray} 
\delta \phi _n&=&\epsilon _{n+1}+\epsilon _n \,\,\,\, (n\geq 1), \,\,\,\,
\delta \phi _0=\epsilon _1\,\, , 
\label{mlett7} \\  
\delta \pi _n&=&-\epsilon _{n+1}'+\epsilon _n' \,\,\,\, (n\geq   1), \,\,\,\,
\delta \pi _0 =-\epsilon _1'\,\, ,  
\label{mlett8}\\
\delta \lambda _n&=&-\dot \epsilon _n+ (-1)^n \epsilon _n' \,\,\,\,
\,\,(n\geq 1)\,\,\,\, , 
\label{mlett9} 
\end{eqnarray} 
\end{mathletters} 
$\epsilon _n\equiv \epsilon _n(x)$.   These  gauge symmetries enable
one to gauge away the variables
($\phi _n, \pi _n$) ($n\geq 1$) added to eliminate the second
class constraints, and leaves one with a single chiral boson.
To see this, 
it is more convenient to replace the pairs ($\phi _n, \pi _n$) by the
self-conjugate variables

\begin{equation}
\mu _n=\pi _n-\phi _n' \label{10}
\end{equation}
and
\begin{equation}
\nu _n= \pi _n+\phi _n' \label{11}
\end{equation}
with brackets
\begin{mathletters}
\label {12}\begin{eqnarray}
\left[\mu _n (\sigma ),\mu _m (\sigma ')\right]&=&
-2 \delta _{nm}
\frac{\partial }{\partial \sigma} \delta (\sigma -\sigma ')
 \,\, , \label{mlett13} \\
\left[\nu _n (\sigma ),\nu _m (\sigma ')\right]&=&2 \delta _{nm}
\frac{\partial }{\partial \sigma} \delta (\sigma -\sigma ')
\, ,  \label{mlett14} \\
\left[\mu _n (\sigma ),\nu _m (\sigma ')\right]&=&
0 \,\,\,\,\, ,\label{mlett15}\end{eqnarray}
\end{mathletters}
in terms of which the constraints and gauge transformations read
\begin{mathletters}
\label{16}\begin{eqnarray}
T_m&=&\mu _{m-1}+\nu _m \,\,\,\, (m\geq 1) \label{mlett17} \\\
\delta \mu _m&=&-2\epsilon _{n+1}'\,\,\, , \,\,\, \delta \nu _n=2
\epsilon _n'\,\,
\, (n\geq 1) \label{mlett18} \\\
\delta \mu _0 &=& -2\epsilon _1 ' \,\,\, , \,\,\, \delta \nu _0 = 0
 \,\,\,\,\, .\label{mlett19}
\end{eqnarray}
\end{mathletters}

On the constraint surface, one may eliminate all the $\mu _n$'s ($n\geq 0$) in
terms of the $\nu _n$'s. Thus, the most general function on the constraint
surface may be assumed 
to depend only on the $\nu _n$'s.
This function will be gauge invariant if and only if it actually does not
 involve the $\nu _m$'s 
for $m\geq 1$, since these
variables transform independently under gauge transformations. Thus, the
more general
gauge invariant function may be assumed to depend only on the single
variable $\nu_0$, which 
is self-conjugate.  This means that the reduced phase space
(see e.g. \cite{Hen}, chapter 2) of the system is indeed that of a single
chiral boson.

Differently put, one may impose the gauge condition $\nu _n=0$ ($n\geq 1$)
to gauge away $\nu _n$. 
Once this is done, the $\mu _n$'s must vanish by the constraints, and 
only the single chiral variable $\nu _0$ 
is left.

\subsection{ Path integral}

Let us now turn to the quantization of the system. We shall adopt the
path integral approach. The most expedient way to get the gauge fixed action
to be path-integrated is to use the antifield formalism \cite{Batil,Hen,Gom}.
The solution of the master equation  for (\ref{1}) is easily constructed
to be
\begin{eqnarray}
S=S_0 + \int d^2x \{\sum _{n\geq 1}\left[ \phi _n^* \left(\sigma _{n+1}+
\sigma _n\right)  \right. \cr+ \pi ^*_n 
\left( -\sigma _{n+1}'+\sigma _n'\right) \cr  +\left. \left.
 \lambda ^* _n\left( 
-\dot 
\sigma _n +(-1)^n \sigma _n'\right) 
\right]+
 \left( \phi _0^*
\sigma _1 + \pi _0^* \sigma _1'\right)\right\},\label{20}
\end{eqnarray}
where the $\sigma _n$ are the ghosts associated with the gauge symmetry
(\ref{6});
$\phi ^*_n \, ,\, \pi _n^*$, $\lambda ^*_n$ and $\sigma ^*_n $ are the
antifields. In order
to fix the gauge one needs to choose an appropriate gauge fixing fermion,
and our
goal is to end up with a covariant path integral. The original action (\ref{1})
is not manifestly covariant because of the $\lambda _n$-terms. If those terms
were absent, we would have an infinite number of uncoupled  scalar fields,
with action equal to the standard, covariant Klein-Gordon action
(in Hamiltonian form). This 
suggests imposing the gauge $\lambda _n=0$. This can be
achieved by interchanging the roles of $\lambda _n $ and $\lambda _n^*$ and 
by taking as gauge fixing fermion $\psi =0$\cite{Hen} (exercise
19.14). One finds then
\begin{mathletters} 
\label{21}\begin{eqnarray}
\lambda _n&=&-\frac {\delta \psi}{\delta \lambda ^*_n}=0 \,\,\, ,\,\,\, 
\phi ^*_n=\frac {\delta \psi}{\delta \phi _n}=0 \\  \pi ^*_n&=&\frac{\delta
\psi }{\delta \pi _n}=0 \, \,\,\,\, ,\,\,\,\, \sigma ^*_n= \frac{\delta \psi
}{\delta \sigma _n}=0 \,\,\, .
\end{eqnarray}
\end{mathletters}
With that gauge choice, the effective action is just
\begin{equation}
S^{\rm eff}= S_0^{(\lambda =0)}+\int d^2x \sum _{n\geq 1} \lambda ^*_n
\left(- \dot \sigma _n
+(-1)^n\sigma _n'\right)\label{22}
\end{equation}

It is straightforward to integrate over the momenta $\pi _n$. One gets
\begin{eqnarray}
S^{{\rm eff}}= \int d^2x \left[\sum _{n=0}^{\infty} \left( (-1)^n {\frac{1}{2}} 
\left( \dot \phi ^2_n-\phi _n^{\acute {2}}  \right) \right) 
\right.\cr \left.  +\sum _{n\geq 1}\left[
\lambda ^*_n\left( -\partial _0 +(-1)^n 
\partial _1\right)\sigma _n\right]\right]\,\,\,\, .\label{23}
\end{eqnarray}
This is the action of \cite{Mc} if one makes the identification $\lambda ^*_n=
\bar C_{n-1}$ and $\sigma _n=b_{n-1}$. This effective action is manifestly
covariant and has been shown in \cite{Mc} to yield a path integral
reproducing the correct amplitudes when properly
handled. The covariance of (\ref{23}) is
manifest if one rewrites it as
\begin{equation}
S^{\rm eff}= \int d^2x \left[\sum _{n=0}^{\infty}  (-1)^n \frac{1}{2} 
\partial _{\mu } \phi \partial ^{\mu }\phi +\sum _{n\geq 1}\lambda
 ^{*\mu }_n\partial _{\mu }\sigma _n\right]\label{24}
\end{equation}
where $\lambda ^{*\mu }_n $ is a vector obeying the covariant algebraic
constraint

\begin{equation}
\lambda ^{*\mu }_n=(-1)^n \eta ^{\mu \nu }\epsilon _{\nu \rho}\lambda ^{\rho}_n
\label{25}
\end{equation}
and having thus only one independent component. The partition function is
\begin{equation}
Z=\int D\phi D\lambda ^* D\sigma \,\, exp [iS^{\rm eff}]\,\,\,\, .\label{26}
\end{equation}

\section{First class formulation of chiral p-forms}

It is possible to generalize the first class formulation of chiral bosons
discussed above to chiral p-forms in ($2p+2$)-dimensions (with p even). The
starting point is the action of \cite{Henteit1,Henteit2}
\begin{eqnarray}
{S}\left[\pi , A, \lambda \right] =\int d^{N}x\left[  
\pi _{(0)}^{j_1...j_p}\dot A_{j_1...j_p}{(0)}\right. \cr \left.-{\cal H}- 
\lambda _{i_1...i_p}^{(0)}
\left( \pi _{(0)}-\beta _{(0)}\right) ^{i_1...i_p}\right]\,\,\,\, ,
\label{27}
\end{eqnarray}
with $N=2p+2$ (equation (78) of \cite{Henteit2}). Here
\begin{equation}
{\cal H}=\frac{1}{2}\left(\pi _{(0)}^2+\beta _{(0)}^2\right)\label{28}
\end{equation}
is the Hamiltonian density in flat space, while $\lambda ^{(0)}_{i_1...i_p}$
is the Lagrange multiplier for the chiral constraint
\begin{equation}
(\pi _{(0)}-\beta _{(0)})^{i_1...i_p}\approx 0  \,\,\, .\label{29}
\end{equation}
In (\ref{27}), $\pi _{(0)}^{i_1...i_p} $ is the momentum conjugate to
$A_{(0)}^{i_1...i_p}$ and $\beta _{(0)}^{i_1...i_p}$ is the ``magnetic
component" of the field strength $F$ explicitly given by

\begin{mathletters} 
\label{30}
\begin{eqnarray} 
 F_{i_1...i_{p+1}}=\partial _{i_1}A_{i_2...i_{p+1}}+p 
 \,\,{\rm cyclic \,\, terms}
\label{mlett31} \\ 
\beta \equiv \beta ^{i_1...i_p}= \frac {1}{(p+1)!} \epsilon ^{i_1...i_{2p+1}}
F_{i_{p+1}...i_{2p+2}}
 \,\,\,\, .
\end{eqnarray}
\end{mathletters}
If one solves the chiral constraint in (\ref{27}), one gets
the action 
$S[A_{(0)}^{i_1...i_p}]=\frac{1}{p!}\int d^Nx\left(\varepsilon _{(0)}
 .\beta _{(0)} -\beta _{(0)} ^2\right)
$ of \cite{Henteit1}, where $\varepsilon _{(0)}$ is the ``%
electric  component" of the field strength $F$.

As pointed out in \cite{Henteit1} the chiral constraint (\ref{29}) is no
longer pure
second class, contrary to what happens in the case p=0.
 Rather, the divergence of (\ref{29}) is first class
\begin{equation}
G_{(0)}^{i_1...i_{p-1}}= \pi _{(0)}^{i_1..i_p},_{i_p}=0 \,\,\, \, .\label{34}
\end{equation}
It is convenient to enlarge the set of constraints by including
explicitly
(\ref{34}). This is permissible since it simply amounts to replace the original
description of the constraint surface by an equivalent (but reducible) one.
In this redudant description the action reads

\begin{eqnarray}
S[\pi _{(0)},A_{(0)},
\lambda _{(0)}]=
 \int d^{N}x\left[
\pi_{(0)}^{j_1...j_p}\dot A^{(0)}_{j_1...j_p}-H\right.\cr \left.
- \lambda _{i_1...i_p}^{(0)}
\left( \pi _{(0)}-\beta _{(0)}\right)^{i_1...i_p}-pA^{(0)}_{0i_1...i_{p-1}}
G^{i_1...i_{p-1}}_{(0)}\right]\,\, , 
\label{35}
\end{eqnarray}
where $A^{(0)}_{0i_1...i_{p-1}}$ is the Lagrange multiplier for ``Gauss law"
(\ref{34}).

The chiral constraint (\ref{29}) has also, as in the chiral boson case, a
second class component.
The first step in reaching a manifestly covariant path integral is to
reformulate the system in such a 
way that there are only first class
constraints. This can be achieved by enlarging the original phase space
\cite{Fad,Bat}. As in the chiral boson case, one needs
an infinite
set of auxiliary variables. We shall not give here 
all the details of the procedure,
but instead, we shall directly give the final answer and check that it
is indeed
correct.

The purely first class formulation of a chiral p-form in ($2p+2$)-dimensions
is given by the action
\begin{equation}
S_0\left[A^{(n)}_{\mu _1...\mu _p},\pi ^{i_1...i_p}_{(n)},
\lambda ^{(n)}_{i_1...i_p}\right]=S^{(2)}_0+S^{(\lambda )}\label{36}
\end{equation}
where $S^{(2)}_0$ is just the action for an infinite number of non-chiral
p-forms, 
in Hamiltonian form,
\begin{eqnarray}
S_0^{(2)}
 =\int d^{N}x\left[\sum ^{\infty}_{n=0}
\pi_{(n)}^{j_1...j_p}\dot A_{j_1...j_p}^{(n)}-{\cal H}\right.\cr
\left.-pA^{(n)}_{0i_1...
i_{p-1}}
G^{i_1...i_{p-1}}_{(n)}\right]\,\,\,\ 
\label{37}
\end{eqnarray}
\begin{equation}
 {\cal H}\approx \frac{1}{2}   \sum ^{\infty } _{n=0} \left[\pi ^2_{(n)}+\beta 
 _{(n)}^2\right]
(-1)^n
\,\,\,\,\, , \label{38}
\end{equation}
while $S^{\lambda}$ is the lagrange multiplier term enforcing
the infinite set of first class constraints $T_{(n)}^{i_1...i_p}=0$,
\begin{equation}
S^{\lambda }= \int d^{N}x\sum ^{\infty}_{n=1}\left(\lambda ^{(n)}_{i_1...i_p} 
T_{(n)}^{i_1...i_p}\right)  
\end{equation}
\begin{eqnarray}
T^{k_1...k_p}_{(n)}=\pi^{k_1...k_p}_{(n)} -\beta^{k_1...k_p}_{(n)}
 +\pi^{k_1...k_p}_{(n)} +\beta^{k_1...k_p}_{(n)} \,\,\,\,\,\, \cr  (n\geq 1) 
 \label{39}
\end{eqnarray}

\begin{equation}
\frac{\delta S_{\lambda }}{\delta \lambda ^{(n)}_{j_1...j_p}}=0 \,\,\,
\Longleftrightarrow
\,\,\,\,T^{j_1...j_p}_{(n)}=0 \,\,\, .\label {40}
\end{equation}
The constraints  $T_{(n)}^{j_1...j_p}=0$ are  easily verified to be first class
among themselves

\begin{equation}
\left[ T_{(n)}^{j_1...j_p},  T_{(m)}^{k_1...k_p}\right]= 0 \,\,\,\, ,
\label {41}
\end{equation}
and to commute weakly with the Hamiltonian $\int {\cal H}d^{N-1}x$. Note
that this holds only because  p  is even.

The first class constraints  $T_{(n)}^{j_1...j_p}=0$ and
$G^{j_1...j_{p-1}}_{(n)}=0$ are
non independent since 
$G_{(m-1)}^{j_1...j_{p-1}}+
G_{(m)}^{j_1...j_{p-1}}\equiv T_{(m)}^{j_1...j_p},_{j_p}$. Furthermore 
$ G_{(n)}^{j_1...j_{p-1}},_{j_{p-1}} \equiv 0$. The constraints 
are clearly highly reducible.

To verify the equivalence of the system (\ref{36}) with the original system
describing a single chiral p-form, one can proceed as 
in the chiral boson case,
taking this time due account of the presence of the gauge freedom caracteristic
of the p-forms, $A^{(n)}_{j_1...j_p} \rightarrow  A^{(n)}_{j_1...j_p} +
(d\epsilon )_{j_1...j_p}^{(n)}$ were $\epsilon $ 
is an arbitrary  (p-1)-form.
Because of that gauge freedom, the observables (``gauge invariant functions")
may be assumed to involve only $\pi _{(n)}^{j_1...j_p}$ and $\beta
 _{(n)}^{j_1...j_p}$.
These variables are, however, not invariant under the gauge transformations
generated   by the $T$'s. To analyse
the implications of this additional invariance,
we make a change of variables analogous to (\ref{10}-\ref{11})
\begin{equation}
\mu ^{j_1...j_p}_{(n)}=\pi _{(n)}^{j_1...j_p}-\beta _{(n)}^{j_1...j_p}
\label{42}
\end{equation}

\begin{equation}
\nu ^{j_1...j_p}_{(n)}=\pi _{(n)}^{j_1...j_p}+\beta _{(n)}^{j_1...j_p}\,\,\,
 .\label{43}
\end{equation}
These variables are not independent since they are both divergenceless
\begin{equation}
\mu ^{j_1...j_p}_{(n)},_{j_p}=0  \,\,\,\,\,\,\,   \nu ^{j_1...j_p}_{(n)}
,_{j_p}=0
\,\,\,\, ,\label{44}
\end{equation}
as it follows from $\beta _{(n)}^{j_1..j_p},_{j_p}\equiv 0$ 
and  the Gauss's law
constraint $G^{j_1...j_{p-1}}_{(n)}\approx 0$. They fullfill the following
bracket relations

\begin{mathletters}
\label{45}\begin{eqnarray}
\left[\mu _{(n)}^{i_1...i_p} (\sigma ),\mu _{(m)}^{j_1...j_p} (\sigma ')
\right]&=&
-2 \delta _{nm}
\epsilon ^{i_1..i_p i_rj_1..j_p} \partial _{i_r}\delta 
 \,\, \\
\left[\nu _{(n)}^{i_1...i_p} (\sigma ),\nu _{(m)}^{j_1...j_p} (\sigma ')
\right]&=&2 \delta _{nm}
\epsilon ^{i_1..i_pi_rj_1..j_p}\partial _{i_r}\delta 
\,\,   \\
\left[\mu _{(n)}^{i_1...i_p} (\sigma ),\nu _{(m)}^{j_1...j_p} (\sigma ')
\right]&=&
0 \,\,\,\,\, \end{eqnarray}
\end{mathletters}
where $\delta \equiv \delta (\sigma -\sigma ')$.

The interest of  the new variables is that they have simple transformations 
properties under the gauge freedom generated by the new constraints
$T_{(n)}$'s,
namely

\begin{mathletters}
\label{46}
\begin{eqnarray}
\delta _{\epsilon }\mu _{(n)} ^{k_1...k_p}&=& 
-2 \epsilon ^{k_1...k_pk_rj_1...j_p}
\partial _{k_r}\epsilon _{(n+1)j_1...j_p} \,\,\, (n\geq 0)\label{mletta}\\
\delta _{\epsilon }\nu _{(n)} ^{k_1...k_p}&=& 
2 \epsilon ^{k_1...k_pk_rj_1...j_p}
\partial  _{k_r}\epsilon _{(n)j_1...j_p} \,\,\,\,\,\, (n\geq 1) \,\,\, , 
\label{mlettb}\\ 
\delta \nu _0 ^{k_1...k_p}&=& 0\,\,\,\,\, .\label{mlettc}
\end{eqnarray}
\end{mathletters}

The constraints $T_{(n)}\equiv \mu _{(n-1)}-\nu _{(n)}=0$ $(n\geq 1)$
enable one to eliminate all the $\mu _{(n)}$'s
($n\geq 0$) in terms 
of the $\nu _{(n)}$'s. Thus the most general function on the constraint
surface,
invariant under the usual p-form gauge symmetries, may be assumed 
to depend only on the
 $\nu _{(n)}$'s, subject to the
transversality condition $\nu _{(n)}^{i_1...i_p},_{i_p}=0$.
This function will be gauge invariant if and only if it actually does not 
involve the $\nu _{(n)} $'s for $n\geq 1$ since the variables can be
 completely gauged
away by the gauge transformations (\ref {46}) (any $\nu _{(n)}^{k_1...k_p}$ 
subject to $\nu _{(n)}^{k_1...k_p},_{k_p}=0$ can be written as
 $\epsilon ^{k_1...k_pi
j_1...j_p}
\partial _i \epsilon _{(n)j_1...j_p}$). Thus, the reduced phase space of the 
system is spanned by the single variable $\nu _{(0)}^{k_1...k_p}(\vec x)$
 obeying
the commutation relations (\ref{45}) and subject to the transversality
condition
(\ref{44}), exactly as in the original description.

A different way to say the same thing  is to observe that a partial
gauge fixing
is given by $\nu ^{k_1...k_p}_{(n)}=0\,\,\,\, (n\geq 1)$ and
$A_{(n)}^{k_1...k_p},
_{k_p}=0\,\,\, (n\geq 1)$. The residual gauge freedom is just the standard
gauge freedom of the 0-th p-form.  One then finds, since the  gauge 
 conditions
 and the
constraints imply
together $A_{j_1...j_p}^{(n)}=0\,\,\, $ and $\pi _{(n)} ^{k_1...k_p}=0\,\,\,
(n\geq 1)$,
that the action (\ref{36}) reduces to the original action (\ref{27}),
establishing again equivalence.

\section{Minimal solution of the master equation}

We now proceed to the construction  of the solution of the master
equation. For definiteness and simplicity of notations, we consider the case
of a chiral 2-form
in 6 dimensions. We shall comment on the general case at the end of the paper.

The equations of motion for the canonical momenta $\pi ^{kl}_{(n)}$ can be
solved to
express them in terms of the 2-form components $A_{\lambda \mu}^{(n)}$ and the
multipliers $\lambda _{km}^{(n)}$. One says that the $\pi ^{kl}_{(m)}$
are ``auxiliary
fields". We shall work from now on with the action $S_0[A^{(n)}_{\lambda \mu},
\lambda
_{km}^{(m)}]$ obtained by eliminating the $\pi$'s using their own equations
of motion,
which is permissible \cite{Hen1}. We shall
not need the explicit form of the action $S_0
\left[ A^{(n)}_{\lambda \mu},\lambda ^{(n)}_{km}\right]$ for all $\lambda$'s.
We shall just need the form of $S_0
\left[ A^{(n)}_{\lambda \mu},\lambda ^{(n)}_{km}\right]$ for $\lambda 
^{(n)}_{km}=0$ because we shall impose this condition when fixing the gauge.
If $\lambda _{km}^{(n)}=0$, the expression
for the
momenta in terms of the $A^{(n)}_{\lambda \mu}$ is
\begin{equation}
\pi _{(m)}^{kl} =- F^{0kl}_{(m)}\,\,\,\, , \label{55}
\end{equation}
as for non-chiral 2-forms. Therefore $S_0$ reduces, 
when $\lambda ^{(n)}_{km}=0$,
to the sum of the
standard
actions for non-chiral p-forms, one for each $A^{(n)}_{\lambda \mu}$,
\begin{eqnarray}
S_0[A^{(n)}_{\lambda \mu}, \lambda ^{(n)}_{km} =0]=\cr - \int
 d^6 x\left[ 
\sum _{n\geq 0}
\left(\frac{1}{6} F^{(n)}_{\lambda \mu \nu} F^{\lambda \mu \nu }_{(n)}\right)
(-1)^n\right]\,\, , \label{56}
\end{eqnarray}
\begin {equation}
F_{\lambda \mu \nu }^{(n)} =3 \partial _{[\lambda } A^{(n)}_{\mu \nu]}\,\,\,\,
. \label{57}
\end{equation}
This action is manifestly covariant.

The action $S_0[A_{\lambda \mu }^{(n)}, \lambda _{km}^{(n)}]$ (for
all $\lambda$'s) is invariant
under the
usual 2-form gauge transformations, which are 
\begin{mathletters}\label{58}
\begin{eqnarray}
\delta _{\epsilon} A^{(n)}_{\mu \nu}&=&\partial_{\mu}\epsilon 
^{(n)}_{\nu}-\partial _{\nu}
\epsilon ^{(n)}_{\mu} \,\,\,\,\,\,\, (n\geq 0) \label{mlett59}\\
\delta _{\epsilon} \lambda ^{(n)}_{km}&=&0 \,\,\,\,\, . \label{mlett60}
\end{eqnarray}
\end{mathletters}
These transformations are generated by  the Gauss constraints $G_{(n)}^{k}
\approx
0$.
The action
is also invariant under the gauge transformations associated with the chirality
constraints 
$T_{(n)}^{kl}\approx 0$, 
which read explicitly, in covariant form,

\begin{mathletters} 
\label{61}
\begin{eqnarray} 
 \delta _u A^{(0)}_{\mu \nu}&=& u_{\mu \nu }^{(1)}
  \label{mlett62} \\   \delta _uA^{(n)}_{\mu \nu}&=&u^{(n)}_{\mu \nu}+
  u^{(n+1)}_{\mu \nu}
\,\,\,\,\,\, (n\geq 1)  \label{mlett63}\\
 \delta _u\lambda ^{(n)}_{kl}&=& -H^{(n)}_{0kl}[u]+\frac{1}{6} (-1)^{(n)} 
 \epsilon _{klpqr}H_{(n)}^{pqr}
[u] \,\,\,\,\,\,\, \,\,\, ,\label{mlett64}
\end{eqnarray} 
\end{mathletters}
where the $H_{\mu \nu \rho }^{(n)} $ are the strength tensor components for
the gauge parameters ($u_{\mu \nu }^{(n)}=-
u_{\nu \mu }^{(n)}\, , \, n\geq 1$)  
\begin{equation}
H^{(n)}_{\mu \nu \rho }[u]=3\partial _{[\mu }u^{(n)}_{\nu \rho ]} \,\,\,\, 
\,\,\,\,\,\,\, .\label{65}
\end{equation} 
The invariance of the action under (\ref{61}) is most easily verified in the
Hamiltonian formalism. If one
takes $u^{(m)}_{0k}=0$, the transformation (\ref{61})
(together with $\delta \pi ^{(n)kl}=\left[ \pi ^{(n)kl},\sum _m  \int d^5x
\left(u^{(n)} _{pq} T^{pq}_{(m)}\right)\right]$ are just the standard gauge
transformations
generated by the
constraints $T_{(m)}^{pq}$, whereas
the transformations (\ref{61}) with $u_{ml}^{(n)}=0$ arise because the
constraints $(
G^k_{(m)},T_{(m)}^{kl})$ are not independent (see \cite{Hen}, chapter 3).
These transformations leave the Hamiltonian action (\ref{36}) invariant and
thus also the action $S_0\left[ A_{\lambda \mu }^{(n)},\lambda ^{(n)}_{km}
\right]$ obtained by eliminating the auxiliary fields $\pi _{(n)}^{kl}$.

The gauge transformations (\ref{58}), (\ref{61}) form a complete set.
However, they are 
not independent. If one takes 
\begin{mathletters}\label{66}
\begin{eqnarray}
u^{(n)}_{\mu \nu}&=& \partial _{\mu } k^{(n)}_{\nu }- \partial _{\nu }
 k^{(n)}_{\mu } \,\,\,\,\,\, 
\,(n\geq 1)\label{mlett67}\\
\epsilon ^{(n)}_{\nu }&=&-k^{(n+1)}_{\nu }-k^{(n)}_{\nu } +\partial _{
\nu }
 \Lambda ^{(n)} \,\,\,\,
\,\,\, (n\geq 1)\label{mlett68}\\
\epsilon ^{(0)}_{\nu }&=&-k^{(1)}_{\nu }+\partial _{\nu }\Lambda ^{(0)}
\,\,\,\,\, ,\end{eqnarray}
\end{mathletters}
one gets  zero field variations for any choice of $k_{\mu }^{(n)} \,\,\,
(n\geq 1)$ and $\Lambda ^{(n)}_{\mu } \,\,\, (n\geq 0)$.

These are the basic ``reducibility identities" and they are not, in turn,
independent. If one takes
\begin{mathletters}\label{69}
\begin{eqnarray}
k^{(n) }_{\nu }&=&\partial _{\nu } \varphi ^{(n) } \label{mlett69}\\
\Lambda ^{(0)} &=&\varphi ^{(1)}\,\,\, , \,\,\,\Lambda ^{(n) }=
\varphi ^{(n+1)}+\varphi ^{(n)} 
\,\,\,\,\,\,\, , (n\geq 1)\label{mlett70}
\end{eqnarray} \end{mathletters}
one gets identically vanishing gauge parameters in  (\ref{66}). There is no
further ``reducibility of the reducibility".

Since the gauge transformation are abelian and the reducibility identities
linear
and holding off shell, the minimal solution of the
master equation is easy to work out.
One gets, following the well-known procedure,

 \begin{eqnarray}
 S^{min} = S_0[A,\lambda ] = \,\,\,\,\,\,\,\,\,\,\,\,\;\;\;\;\;\;  \cr 
 +\int d^6x \left\{\sum _{n\geq 0}\left[
  {A^*}_{(n)}^{\mu \nu } 
\left( \partial _{\mu }C^{(n)}_{\nu }-
 \partial
 _{\nu }C_{\mu }^{(n)}\right) \right. \right. \cr + 
\left. C^{*\nu }_{(n)} \partial _{\nu }
\rho ^{(n)} \right] \cr
+\sum _{n \geq 1}\left[{A^*}^{\mu \nu}_{(n)} \left( \eta ^{(n)}_{\mu \nu}
 +\eta ^{(n+1)} _{\mu \nu}
 \right)  -{\lambda ^*}_{(n)}^{kl} \left( \bar B _{kl}(\eta )^{(n)}\right)
  \right. \cr -{C^*}^{\nu}_{(n)}
\left( \sigma ^{(n)}_{\nu }+ \sigma ^{(n+1)}_{\nu }\right)  +
  {\eta ^*}^{\mu \nu} \left( \partial _{\mu} \sigma ^{(n)}_{\nu }-
   \partial _{\nu}
\sigma ^{(n)}_{\mu }\right) 
 \cr \left. +{\sigma ^*}_{(n)}^{\nu }\partial _{\nu } \chi ^{(n)}
 +
{\rho ^*}^{(n)} \left( \chi ^{(n)} +\chi ^{(n+1)}\right) \right] \cr+ \left.
A^{*\mu \nu }_{(0)}\eta ^{(1)}_{\mu \nu }-C^{*\nu}_{(0)}\sigma^{(1)}_{\nu }
+\rho _{(0)}^*\chi ^{(1)}
\right\}\label{71} \end{eqnarray}
The ghosts $C^{(n)}_{\nu }\,\, (n\geq 0)$ are associated with the 2-form
gauge symmetry (\ref{58}) and have ghost number one. 
Their antifields are $C^{*\nu }_{(n)}$
and have ghost number -2. The
ghosts $\eta ^{\mu \nu }_{(n)}\,\, (n\geq 1)$ are associated with
the gauge symmetry (\ref{61}) and have also ghost number one. Their
antifields $\eta^{* \mu \nu}_{n}$ have ghost number -2.
We have defined $\Theta
^{(n)}_{\rho \sigma} (\eta )$ to be the field strengths of the $\eta $'s

\begin{equation}
\Theta ^{(n)}_{\rho \sigma \mu}= 3\partial _{[\rho }\eta ^{(n)}_{\sigma \mu ]}
\label{72}
\end{equation}
and
\begin{equation}
\bar B_{kl}^{(n)}=\Theta ^{(n)}_{0kl}-\frac{1}{6}\epsilon _{klpqr} 
\Theta ^{(n)pqr}(-1)^n
\,\,\,\, .\label{74}
\end{equation}

Finally we have the following ghosts of ghosts and antifields corresponding
 to the
various reducibilities
\begin{mathletters}
\label{75}
\begin{eqnarray}
\rho ^{(n)}\, ,\,\,\,\, gh\rho ^{(n)}&=&2 , \,\,\,\,\rho ^{*(n)} , 
\,\,\,\,gh \rho ^{*(n)}=-3\,\,\, ,
n\geq 0 \label{mlett76}\\
\sigma ^{(n)}_{\nu }\, , \,\,\,\,gh \sigma _{\nu }^{(n)} &=&2 ,\,\,\,\, 
\sigma ^{*\nu}_{(n)}
\, , \,\,\,\, gh \sigma ^{*\nu }_{(n)}=-3\,\,\, , n\geq 1\label{mlett77}\\
\chi ^{(n)}\, , \,\,\,\, gh \chi ^{(n)} &=&3 \, ,\,\,\,\, 
\chi ^{*}_{(n)}
\, ,\,\,\,\, gh \chi ^{*} _{(n)}=-4\,\,\,\, , n\geq 1\,\, .\label{mlett78}
\end{eqnarray}
\end{mathletters}

\section{Temporal gauge}

One can verify the correctness of the minimal solution
of the master equation by writing the path integral in the ``temporal gauge"
$A_{k0}^{(n)}=0, \,  C^{(n)}_0=0  \,\, (n\geq 0), \,\,  \lambda 
_{(n)}^{kl}=0 , \,\, 
\eta ^{(n)}_{k0}=0 , \,\,\, \sigma _0^{(n)}=0\,\, (n\geq 1)$. This gauge
 fixing can be
 reached
without need for non minimal variables, by exchanging the roles of the  fields
that are set equal to zero for their antifields, and by taking $\psi =0$ 
\cite{Hen}. The antifields 
conjugate to the fields  $A_{k0}^{(n)} ,\,  C_{0}^{(n)}, \lambda
_{k0}^{(n)}, \eta _{k0}^{(n)} $and $\sigma _{0}^{(n)}$  play 
the role of antighosts and will
 be denoted in the
remainder of this section as
\begin{equation}
A^{*0k}_{(n)}\equiv \bar C^{k}_{(n)},  C_{(n)}^{*0}\equiv \bar 
C_{(n)}
, \eta ^{*0k}_{(n)}\equiv \bar \mu ^k_{(n)}  , \sigma _0^{*(n)}
\equiv \bar \mu ^{(n)} 
.\label {79}
\end{equation}

The partition function is then
\begin{eqnarray}
Z=\int DA^{(n)}_{kl} DC^{(n)}_k D\rho ^{(n)} D\bar C_k^{(n)}
D\bar C^{(n)} D\lambda ^{*kl}_{(n)}\cr \times D
\eta _{kl}^{(n)}D 
\sigma _k^{(n) }D\chi ^{(n)}
 D\bar \mu ^{(n)}_k D\bar \mu ^{(n)} exp [iS^{\rm eff}_{tem}]\label{80}
\end{eqnarray}
with
\begin{eqnarray}
&S&^{\rm eff}_{tem}=\int d^6x\left[ \sum _{n\geq 0} \left(-\frac{1}{2}
 F_{0kl}^{(n)} F_{(n)}^{0kl}-\frac 
{1}{6} F_{klm}^{(n)} F_{(n)}^{klm}
\right. \right. \cr &+& \left. \bar C^{k}_{(n)} \partial _0 C_{k}^{(n)}+\bar 
C^{(n)}\partial _0
\rho _{(n)}\right)
\cr &-& \left. \sum _{n\geq 1}\left( \lambda ^{*kl}_{(n)}\bar B^{(n)}
\eta ^{(n)}_{(kl)}- 
\bar \mu ^{k}_{(n)}\partial _0\sigma _k^{(n)}-\bar \mu ^{(n)} \partial _0
\chi _{(n)} \right) \right]\label{81}
\end{eqnarray}

The partition function (\ref{80}) is equal to an infinite
 product of determinants which can be evaluated as
 follows.
The second order differential operator $D$ acting on the $A_{kl}$'s in the
 Euler-Lagrange equations 
following from the gauge fixed action (\ref{81})  can be written
as a product
of first order differential operators,
\begin{mathletters}
\label{82}\begin{eqnarray}
D&=&D_+D_- \label{mlett83}\\ D_+A_{kl}^{(n)}&=& \partial _0 A_{kl}^{(n)}
+\frac{1}{2}\epsilon _{klmrp}
\partial ^m A^{(n)rp}\\
D_- A_{kl}^{(n)} &=&\partial _0A_{kl}^{(n)}-\frac{1}{2}\epsilon _{klmrp}
\partial ^mA^{(n)rp}\label{mlett84}
\,\,\, .\end{eqnarray}
\end{mathletters}

If $A^{rp}_{(n)}$ is ``longitudinal" ($\partial ^{[m} A^{rp]}_{(n)}=0$),
 the operators $D_+$ and
$D_-$ reduce to $\partial _0$, while $D$ becomes $\partial _0^2$. There are 4
longitudinal modes  among the 10 $A_{kl}$'s, and 6 modes transverse to them.
If one
denotes  by $\tilde D_+$ and $\tilde D_-$ the operators induced in the
transverse
subspace, one has formally 
\begin{equation}
det D_+=det \tilde D_+(det \partial _0)^4\, , 
det D_-=det \tilde D_- (det \partial
_0)^4 .\label{85}
\end{equation}
Thus, $det D =det \tilde D_+ det \tilde D_- (det \partial _0)^8$ and
the integration
over $A_{kl}^{(n)}$ yields for each n the factor
\begin{equation}
(det D)^{-\frac{1}{2}}=(det \tilde D_+)^{-\frac{1}{2}}
(det \tilde D_-)^{-\frac{1}{2}}(det \partial _0)^{-4}
 \,\,\,\, .\label{86}
\end{equation}
The integration over the 5 anticommuting ghost pairs $C^{(n)}_{k}$ and
$\bar C^{(n)}_{k}$ ($n$ fixed) clearly yields
$(det \partial _0)^5$, while the integration over the single commuting
 ghost pair $\bar C^{(n)}$ and $\rho ^{(n)}$
gives $(det \partial _0)^{-1}$. 
Accordingly, the integration over ($ A_{kl}^{(n)}, \bar C_k^{(n)}, C^{(n)}_k,
 \rho ^{(n)}, \bar C^{(n)}$) yields, for each given $n$,
\begin{equation}
 (det \tilde D_+)^{-\frac{1}{2}} (det \tilde D_-)^{-\frac{1}{2}}
\,\,\,\, .\label{87}\end{equation}

Consider now the integration over the sector ($\lambda ^{*kl}_{(1)}, 
\eta _{kl}^{(1)},
 \bar \mu _{(1)}^{k}, \sigma ^{(1)}_{k}, \bar \mu ^{(1)}, \chi ^{(1)}$).
The $\lambda ^* \eta $
term can be written as $-\lambda ^{*}_{(1)}D_+\eta _{(1)}$, thus we get from
$\int D\lambda ^*_{(1)}D\eta _{(1)} $ the determinant

\begin{equation}
det D_+=det \tilde D_+ (det \partial _0 )^4 \,\,\,\, .
\label{88} \end{equation}

At the same time, the integration over the commuting ghost pairs
( $\bar \mu _{(1)}^{k}
, \sigma ^{(1)}_{k}$) brings in $(det \partial _0)^{-5}$ and the integration
over the
anticommuting ghost pair  ($\bar \mu ^{(1)} , \chi ^{(1)}$) gives ($ det
\partial _0)^{1}$. Accordingly, the integration over ($ \lambda ^{*kl}_{(1)},
\eta ^{(1)}_{kl}, \bar \mu _{(1)}^{k} , \sigma ^{(1)}_{k}, \bar \mu ^{(1)},
\chi
 ^{(1)}$)  brings in the factor
$(det\tilde  D_+)$. The same argument applies to the  integration for
the other indices $n$ with $n$ odd, while
for $n$ even one gets $(det \tilde D_-)$.

Putting things together, one finds that the partition function Z is equal to
the infinite product 

\begin{eqnarray}
(det \tilde D_-)^{-\frac{1}{2}} (det \tilde D_+)^{-\frac{1}{2}}
 det \tilde
 D_+ (det \tilde D_+)^{-\frac{1}{2}} \cr \times (det \tilde D _-)^{-
\frac{1}{2} }det
 \tilde D_-
 (det \tilde D_-)^{-\frac{1}{2}}....\label{89}
\end{eqnarray}
The
first two factors $(det \tilde D_-)^{-\frac{1}{2}}$ and
$(det \tilde D_+)^{-\frac{1}{2}}$
come from the integration over $A_{(0)}$ and its companion variables, the next
factor $det \tilde D_+$ come from the integration  over $\lambda ^*_{(1)}$
and its companion variables, the next two factors 
$(det \tilde D_+)^{-\frac{1}{2}}$
and  $(det \tilde D_-)^{-\frac{1}{2}}$  come 
from the integration over $A_{(1)}$ and
its companion variables etc...
In order to regularize the expression (\ref{89}), we regroup the factors
 along the ideas of \cite{Mc}
(formula (4.11)),
which follows the way the extra variables have been progressively added. 
More precisely, we rewrite  (\ref{89}) as
 \begin{eqnarray} 
 (det \tilde D_-)^{-\frac{1}{2}}\left( (det \tilde D_+)^{-\frac{1}{2}}
  det \tilde
 D_+ (det \tilde  D_+)^{-\frac{1}{2}}\right)\cr \times \left( (det 
\tilde D _-)^{-\frac{1}{2} }det 
 \tilde D_-
 (det \tilde D_-)^{-\frac{1}{2}}\right)....\label{AAA}
\end {eqnarray}
By regrouping the factors in this manner,  
one finds that the partition function reduces to

\begin{mathletters}
\label{90} \begin{eqnarray}
 Z&=&(det \tilde D_-)^{-\frac{1}{2}}.1.1.1....\\
&=& (det \tilde D_-)\,\,\,\,\, 
 \end{eqnarray}
 \end{mathletters} 
as it should.

\section{Covariant path integral}

While the temporal gauge $A^{(n)}_{0k}=0 \,\,\, , \,\, \lambda _{kl}^{(n)}=0$
does not lead to a manifestly Lorentz invariant effective action, one may
devise gauge conditions that do achieve this goal. For instance, one may 
impose the Lorentz gauge

\begin{equation}
\partial ^{\mu } A^{(n)}_{\mu \nu}=0 \,\,\,\,\, (n\geq 0) \label{100}
\end{equation}
for the ordinary 2-form gauge symmetries, together with

\begin{equation}
\lambda ^{(n)}_{kl}=0 \,\,\,\,\, (n\geq 1) \label{101}
\end{equation}
for the gauge transformation arising from the introduction of the auxiliary
variables. This second condition is intended to eliminate the 
non-covariant Lagrange multiplier term $\sum \lambda _{kl}^{(n)}T^{kl}_{(n)}$
from the action, as in the chiral boson case. 
The gauge conditions (\ref {100})-%
 (\ref{101}) must be supplemented by conditions that freeze the ``ghost
gauge freedom" associated with the reducibility identities, e.g., one may
take
\begin{equation}
\partial ^{\mu }C_{\mu }^{(n)}=0 \,\,\,\, 
(n\geq 0)\, , \,\,\,\,\, \partial^\mu
\eta _{\mu \nu }^{(n)} =0 \,\,\,\, (n\geq 1)\label{102}
\end{equation} 
and

\begin{equation}
\partial ^{\mu } \sigma _{\mu }^{(n)} =0 \,\,\,\, (n\geq 1)\,\,\,\, .
\label{103}
\end{equation}

The gauge condition  $\lambda _{kl}^{(n)}=0$ does not require the introduction
of non minimal variables. It can again be implemented by exchanging the roles
of $\lambda ^{(n)}_{kl}$ and $\lambda ^{*kl}_{(n)}$ and by taking a gauge
fixing
fermion $\psi$  that does not depend on $\lambda ^{*kl}_{(n)}$, so that 
$\lambda _{kl}^{(n)}=-\frac{\delta \psi}{\delta \lambda ^{*kl}_{(n)}}$ indeed
vanishes.

By contrast, the gauge conditions (\ref{100}),(\ref{101}),
(\ref{102}) and (\ref{103}) do
need a non-minimal sector. The non-minimal sector required to the 2-form
gauge symmetry is well known (~\cite{Batil}, \cite{Hen} chapter 19) and
is given by the antighosts $\bar C^{\mu }_{(n)}$, $\bar C^{(n)}$ together
with the auxiliary variables $b ^{(n)}_{\mu },\,\, b^{(n)},\,\, \pi ^{(n)}$
and $\eta ^{(n)}$, with ghost number assignments

\begin{mathletters}
\label{104}
\begin{eqnarray}
gh \bar C^{\mu }_{(n)}   &=&-1 ,\,\,\,\,  gh \bar C ^{*(n)}_{\mu }   =0\\
gh \bar C_{(n)}   &=&-2 ,\,\,\,\,  gh \bar C ^{*(n)}   =1\\
gh b_{\mu }^{(n)}   &=&0, \,\,\,\,  gh b_{(n)}^{\mu }   =-1\\
gh b^{(n)}   &=&-1, \,\,\,\,  gh b_{(n)}^{* }  = 0\\
gh  \pi _{(n)}   &=&1 ,\,\,\,\,  gh \pi ^{*}_{(n)}   =-2\\
gh \eta ^{(n)}   &=&0 ,\,\,\,\,  gh \eta ^{*}_{(n) }   =-1\,\,\,\, .
\end{eqnarray}
\end{mathletters}
The non-minimal term in the solution of the master equation required for
freezing  covariantly $A_{\mu \nu}^{(n)}\rightarrow A_{\mu \nu}^{(n)}+
\partial _{\mu }\Lambda _{\nu }-\partial _{\nu }\Lambda _{\mu }$
is then
\begin{equation}
\int d^6 x \sum _{n=0}^{\infty} \left(\bar C^{*(n)}_{\mu }b^{\mu}_{(n)}+
\bar C_{(n)}^* b_{(n)}+\eta ^*_{(n)}\pi _{(n)}\right)\,\,\, .\label{105}
\end{equation}

Since the gauge conditions and the structure of the minimal solution of
the master equation for the ghost variables $\eta ^{(n)}_{\mu \nu}$ is
quite similar to that for $A^{(n)}_{\mu \nu}$ with mere shift in the ghost
number, we also add to $S$ similar non-minimal terms for imposing 
the conditions $\partial ^{\mu } \eta _{\mu \nu}^{(n)}=0,\,\,\,\,\, \partial
^{\mu }\sigma _{\mu }^{(n)}=0,$

\begin{equation}
\int d^6 x\sum ^{\infty}_{n=1}\left(\bar \sigma ^{*(n)}_{\nu } d ^{\mu }_{(n)}
+\bar \sigma ^{*(n)} d _{(n)}+\mu ^{*}_{(n)} \theta _{(n)}\right)\label{106}
\end{equation}
with

\begin{mathletters}
\label{107}
\begin{eqnarray}
gh \bar \sigma ^{\mu }_{(n)}   &=&-2 ,\,\,\,\,  gh \bar \sigma
 ^{*(n)}_{\mu }   =1\\
gh \bar \sigma_{(n)}   &=&-3 ,\,\,\,\,  gh \bar \sigma ^{*(n)}   =2\\
gh d^{\mu }_{(n)}   &=&-1 ,\,\,\,\,  gh d^{(n)}_{\mu }   =0\\
gh d^{(n)}   &=&-2, ,\,\,\,\,  gh d_{(n)}^{* }  = 1\\
gh  \theta _{(n)}   &=&2 ,\,\,\,\,  gh \theta ^{*}_{(n)}   =-3\\
gh \mu ^{(n)}   &=&1 ,\,\,\,\,  gh \mu ^{*}_{(n) }   =-2\,\,\,\, .
\end{eqnarray}
\end{mathletters}

The complete, non-minimal solution of the master equation appropriate to
the problem at hand is thus

\begin{eqnarray}
 S= S^{\rm min} \,\,\,\,\,\,\,\,\,\,\,\,\,\,\,\,\,\,\,\,\,\,\, \cr 
+\int d^6x\left[\sum _{n=0}^{\infty} \left( 
\bar {C^*}^{(n)}_{\mu} b^{\mu }_{(n)}+
{\bar C^*}_{(n)} b_{(n)} +{\eta ^*}_{(n)} \pi _{(n)}\right)\right. \cr +\sum 
_{n=1}^{\infty}\left(  
 {\bar \sigma ^{*(n)}}_{\nu} d^{\mu }_{(n)}  
 \left. +{\bar \sigma ^*}_{(n)} d^{(n)} +
 {\mu ^*}_{(n)} \theta ^{(n)}\right)\right]
\,\,\, .\label{179} \end{eqnarray}

The final step is to choose a gauge fixing fermion and to 
eliminate
the antifields through the expressions
\begin{equation}
\zeta ^*_{(n)}=\frac{\partial \Psi}{\partial \zeta _{(n)}} \,\,\,\,\,\,\,
 \lambda _{(n)}
=-\frac {\partial
\Psi}{\partial \lambda ^*_{(n)}}\label{180}
\end{equation}
where $\zeta _{(n)}$ is any field or ghost present
in the gauge fixing fermion $\Psi $ (but $\lambda _{(n)}$)
and $\zeta ^*_{(n)}$ the corresponding antifield.
The appropriate gauge fixing fermion that enforces the gauge conditions
(\ref{100}), (\ref{101}), (\ref{102}) and (\ref{103}) is 
\begin{eqnarray}
\Psi = \int d^6 x  \sum _{n=0}^{\infty}\left[ 
\bar C^{\mu }_{(n)}\left(\partial ^{\nu } A^{(n)} _{\mu \nu}\right)
 +\bar C_{(n)} 
\partial ^{\nu } C^{(n)}_{\nu } \right. \cr 
+\bar C^{\nu }_{(n)} \partial _{\nu }
 \eta ^{(n)}
\cr
+\left.\bar \sigma _{(n)}^{\mu } \left(
\partial ^{\nu } \eta ^{(n)}_{\mu \nu }\right)+\bar 
\sigma _{(n)}\partial ^{\nu }\sigma
^{(n)}_{\nu } +\bar \sigma _{(n)}^{\nu } \partial _{\mu } \mu ^{(n)} \right]
\label{181}\end{eqnarray}
(see \cite{Batil}, \cite{Hen} chapter 9)%
. So the final expression for the solution of the master equation is, 
taking 
(\ref{58}) into account,

\begin{eqnarray}
S_{\Psi}= \int d^6 x \left\{\sum ^{\infty }_{n=0} 
\left[-\frac{1}{6}F^2_{(n)}(-1)^n
\right.\right. \cr +\frac{1}{2}\left(\partial ^{\mu }
\bar C^{\nu }_{(n)}-\partial 
^{\nu} \bar C^{\mu }_{(n)}\right)
\left(\partial _{\mu } C_{\nu }^{(n)}-\partial _{\nu } C_{\mu}^{(n)}
\right)\cr
+ \partial_\mu \eta^{(n)} b^\mu_{(n)}-\partial ^{\nu }
\bar C_{(n)} \partial _{\nu} \rho ^{(n)}+
\partial ^{\nu}
A^{(n)}_{\mu \nu}b^{\mu}_{(n)}\cr \left.+
\partial ^{\nu }C^{(n)}_{\nu }b_{(n)}
-\partial _\nu \bar C_{(n)}^{\nu } \pi ^{(n)}\right] 
\cr +\sum_{n\geq 1}
\left[-\lambda ^{*kl}_{(n)}\left(\bar B\right)^{(n)}_{kl}\right. \cr
+ \frac{1}{2} \left(\partial ^{\mu } 
\bar C_{(n)}^{\nu }-\partial ^{\nu }\bar C_{(n)}
^{\mu }\right)
\left(\eta _{\mu \nu }^{(n)}+ \eta _{\mu \nu }^{(n+1)}\right)
\cr
+\partial ^{\nu} \bar C_{(n)}
\left(\sigma ^{(n)}_{\nu}
+\sigma ^{(n+1)}_{\nu }\right)
\cr
+\frac{1}{2}\left( \partial ^{\mu } \bar 
\sigma ^{\nu}_{(n)}
-\partial ^{\nu } \bar \sigma ^{\mu}_{(n)}\right)
\left(\partial _{\mu}\sigma ^{(n)}_{\nu} -\partial _{\nu}\sigma _{\mu}^{(n)}
\right)
 \cr -
\partial ^{\nu} \bar \sigma _{(n)} \partial _{\nu} \chi ^{(n)}+ 
\partial ^{\nu }
\eta ^{(n)}_{\mu \nu}
d^{\mu}_{(n)}\cr +  \partial _{\mu }\mu ^{(n)} d^{\mu}_{(n)} +
\partial ^{\nu} \sigma ^{(n)}_{\nu}
d_{(n)}  -\partial _{\nu} \bar \sigma ^{\nu}_{(n)} \theta _{(n)}
\cr \left.
+ \frac{1}{2}\left( \partial ^{\mu }
\bar C^{\nu }_{(0)}-\partial ^{\nu }\bar C_{(0)}^{\mu }\right)
\eta _{\mu \nu }^{(1)}\right]\cr \left. 
+\partial ^{\nu }\bar C_{(0)} \sigma ^{(1)} _{\nu }\right\}
\label{182}\end{eqnarray}
The action is completely gauge fixed, as one 
easily verifies. All the terms are
 manifestly covariant, including the
term $\lambda ^{*kl}_{(n)} \bar B^{(n)}_{kl}$, which can be written as 
$\tau _{(n)}^{\rho \sigma \lambda } \bar \Theta _{\rho 
\sigma \lambda }^{(n)}$ where the three-rank antisymmetric tensor
$\tau^{\rho \sigma \lambda}$ is  subject to the algebraic constraint

\begin{equation}
\tau_{(n)} ^{\rho \sigma \lambda }=\frac{1}{6}(-1)^n
\epsilon ^{\rho \sigma \lambda 
\alpha \beta \gamma} \tau _{(n) \alpha \beta \gamma},
\label{184}
\end{equation}
which reduces its number of independent components to the 10 independent 
$\lambda ^{*kl}$. [Recall that the $\Theta$'s are the field strengths of 
the ghosts $\eta _{\mu \nu}$, formula (\ref{72}).]

Finally, the same analysis can be repeated along identical 
lines for higher rank chiral p-forms in $2p+2$ 
dimensions ($p=2k, k\geq 1$). One simply needs more
ghosts of ghosts. The procedure follows the standard pattern of the antifield
formalism. The details are left to the reader.

\section{conclusions and prospects}

In this paper, we have obtained a manifestly Lorentz invariant path integral
for a
chiral p-form in (2p+2)-dimensional Minkowskian space-time. Our approach
generalizes
the calculations of McClain, Wu and Yu\cite{Mc} performed for chiral bosons.
The
generalization presents new non-trivial features because the gauge symmetries
are
now reducible. The gauge symmetries that enable one to gauge away the
auxiliary fields
necessary for replacing the second class constraints by first class ones
(leading to
the standard covariant two-point functions) are not 
independent from the standard
p-form gauge symmetries, which are themselves 
already reducible. The correct handling
of this
difficulty requires ghosts of ghosts, absent in the 0-form case, and is most
easily
carried out in the framework of the antifield formalism.

One of the striking features of the manifestly covariant formulation is that
it envolves
an infinite number of auxiliary field variables, as in the chiral boson
treatment. Of
course, the manipulation of an infinite number of variables can be tricky and
even misleading in some calculations, as the attempts to 
derive a manifestly covariant
formulation
of the superparticle through the introduction of an infinite number of
auxiliary
variables have shown \cite{Lind,Fis}. A prescription must be given on how
to compute with
the infinite number of variables . For instance, the terms in
the  infinite sums or
infinite products  that  arise should be grouped in a manner compatible
with the
actual
way the new variables have been progressibly added in order to reach the
covariant
formulation, as  in formula  (\ref {AAA}) above.
More covariant regularizations may be desirable, however. Let us briefly
comment on the gravitational anomaly in this context.

The advantage  of the manifestly covariant formulation  is that 
it enables a
direct coupling to gravity  along the standard lines of ordinary tensor
calculus. The
coupling to gravity in the original non-manifestly covariant formulation
has been
actually worked out first in \cite{Henteit1} (see also \cite{Nie} and \cite
{Bra}), but it does not follow the familiar pattern.

Now, all the terms in the final gauge fixed action 
written in an arbitrary covariant background are
chirally invariant, except the terms
$ \sum _n\lambda  ^{*kl}_{(n)}\tilde B ^{(n)}_{kl}$.
These terms are the only sources of the  
gravitational anomaly. 
Let us denote by $A$ the anomaly due to a single chiral 2-form 
(as evaluated
in \cite{Alv} and \cite{Nie}). The term $\lambda ^{*kl} _1 \bar B_{kl}^{(1)}
(\eta )$ describes a pair of chiral 2-forms of chirality opposite to that of
the original physical chiral 2-form $A^{(0)}_{\mu \nu }$, 
but since these 2-forms are both
fermionic, they are expected to contribute +2A (with the same sign as 
$A^{(0)}_{\mu \nu }$) to the gravitational anomaly. The next fermionic form
$\lambda ^{*kl}_{(2)},\,\, \eta ^{(2)}_{\mu \nu }$ has the same chirality 
as $A^{(0)}_{\mu \nu }$ and contributes -2A. Going on in the same fashion for
higher$ n$'s  one finds that the total contributon  (due to the infinite
number of $(\lambda , \eta )$ pairs)
to the anomaly is given by the infinite sum
\begin {equation}
A' =2A(1-1+1-1...) \label{200}
\end{equation}

This sum is equal to $A$ if one regularizes it as  
$lim _{k\rightarrow -1} (1+k+k^2
+...) =\frac{1}{2}$ . We have not attempted to justify 
this particular regularization
in the present framework but we believe that the above heuristic
derivation indicates the potential usefulness of our approach.

It is hoped to return to this question in the future. It is
 also hoped to analyse in detail the BRST cohomology 
 and the physical spectrum in the covariant formulation.

\vspace{1cm}

{\bf Acknowledgements}

\vspace{0.5cm}
This work has been supported in part by Research contracts from the 
Commission of the European Community, by research 
funds from F.N.R.S. (Belgium) and by CNPq
(Brazil).

\end{document}